\documentclass[12pt,emulateapj]{aastex}
\usepackage{amsmath}
\usepackage{mathtools}
\usepackage{ulem}
\usepackage{natbib}
\usepackage[usenames,dvipsnames]{color}

\newcommand{\Msol}{\mbox{$\rm{M_{\odot}\ }$}}
\newcommand{\msol}{\mbox{$\rm{M_{\odot}\ }$}}

\renewcommand{\emph}{\it}

\begin{document}
\title{
The Impact of Stellar Abundance Variations on Stellar Habitable Zone Evolution
}
\shorttitle{Abundance Variations and Habitable Zones}

\author{Patrick A. Young\altaffilmark{1}, Kelley Liebst\altaffilmark{1}, Michael Pagano\altaffilmark{1}}

\altaffiltext{1}{School of Earth and Space Exploration, Arizona State 
University, Tempe, AZ 85287}

\keywords{stars: abundances --- stars: evolution }

\begin{abstract}
The high quality spectra required for radial velocity planet searches are well-suited to providing abundances for a wide array of elements in large samples of stars. Abundance ratios of the most common elements relative to Fe are observed to vary by more than a factor of two in planet host candidates. This level of variation has a substantial impact on the evolution of the host star and the extent of its habitable zone. We present stellar models of 1\Msol stars with custom compositions representing the full range of these non-solar abundance ratios. We find that the effects derived from variation over the observed range of [O/Fe] has a particularly dramatic effect. Habitability lifetimes for some classes of orbits can vary by gigayears for the observed range in [O/Fe]. 
\end{abstract}

\section{Introduction \label{s:intro}}
In general stellar models are created for different total metallicities scaled relative to the solar abundance pattern, but variations in the abundance ratios at a given [Fe/H] are not considered, except for uniform enhancement of the $\alpha$ elements (O, Ne, Mg, Si, S, Ar, Ca, Ti) at very low metallicities. This is an important problem in its own right, but recent developments in the field of extrasolar planets have added another layer of interest.

Recent discoveries by Kepler of Earth-sized and smaller planets \citep{bor11,charp11,fres12} and planets near or within the nominal habitable zones of stars (i.e. Kepler-22 b \citep{bor12a}, GL 581 c and d \citep{udry07, mayor09} GJ 667C c and d \citep{delfosse12, ang12}) lends a sense of immediacy to the question of what makes a planet habitable . The present position of a star's habitable zone \citep{selsis07,lammer09} can be estimated directly from observations and appropriate assumptions about planetary atmospheres \citep[see][for atmospheric modeling of GL 581 d]{word11, hu11, kalt11}.  The evolution of habitable zones over time is also of fundamental importance. Although life arose quickly on Earth, other steps of critical astrobiological interest took considerably longer. Potentially detectable atmospheric changes produced by the biota, in particular the presence of free oxygen, methane, or nitrous oxide, occurred after roughly a billion years. Complex life 
required a further three billion years. 
Arguments have been proposed that suggest certain evolutionary breakthroughs are low probability events, making it likely that evolutionary timescales will be long as a general rule. \citep{line02, carter08}. The variation of stellar lifetimes with mass and [Fe/H] is a discipline of long standing \citep[][for a recent example]{mowlavi12}, and changes in habitable zone extents with mass and [Fe/H] have been considered \citep[e.g.][]{muir12}, but other abundances may have a substantial effect.

High quality spectra required for radial velocity planet searches are well-suited for determining stellar abundances. For large samples of stars it is possible to statistically separate the intrinsic variation of elemental abundance ratios from observational uncertainties  \citep{desilva06, so09}. An analysis of published abundances for thirteen elements derived from 
the AAT radial velocity planet search \citep{bond06, bond08} provides measurements of the range of variation of abundance ratios relative to Fe for a hundred stars. The sample consists of mostly G and some F and K dwarfs with ages $>$ 3 Gyr and -0.5$<$[Fe/H]$<$0.3, providing an observationally uniform sample that approximates as well as possible the primordial stellar abundances. The elemental ratios relative to the sun ([X/Fe]) vary substantially, more than can be accounted for by
observational error.  The 3$u_{intrinsic}$ variation (this
spread is an rms error, not a standard deviation for a normal
distribution) is close to a factor of two about the mean for C, O, Na, Mg, Al, Si, Ca, and Ti, from 1.7 for Na to 2.3 for O.   Si, Cr, and Ni have
smaller spreads of a factor of 1.4-1.5 about the mean. The r- and s-process elements (species significantly heavier than Fe produced in evolved stars and supernovae) have even larger variations. Also, while these ranges encompass the majority of nearby planet host candidates, there are even more extreme outlier systems \citep{bond08}. The star HD53705 has [O/Fe] = 4, more than 5$u_{intrinsic}$ higher than the mean. HD136352 is similarly enhanced in [C/Fe] \citep{bond06}. Both of these are low metallicity systems ([Fe/H = -0.18 and -0.26, respectively). Enhancements of this magnitude will substantially increase the range of orbits that remain habitable for a multi-Gyr window. 

Stellar evolution calculations are sensitive to the assumed
composition.  This sensitivity arises from two composition-dependent
effects: the equation of state (EOS) and the radiation opacity
\citep{rogers_1996_aa, iglesias_1996_aa}. The equation of state is primarily affected by the number of free particles (less when more nucleons are in heavy nuclei) and the effects of electron screening, where the mutual repulsion by the positive charge of high Z nuclei is reduced by intervening electrons. The changes in the EOS 
are minor, and much less important than the opacity. The elements C, N, O, Ne, Si, Ti, and Fe-peak (stable elements with atomic numbers close to Fe)
elements are particularly important because of their relative
abundance and number of electron transitions. Rearranging the
proportions of different species at a
constant metallicity can result in opacity changes of tens of percent
\citep{iglesias_1996_aa}. Opacity changes the rate of leakage of radiation. 
Increased radiation pressure in the stellar envelope drives expansion, resulting in larger radii and lower effective temperatures. Also, a slower rate of energy loss requires a slower rate of nuclear burning to maintain hydrostatic equilibrium. So we expect stars with enhanced abundances to be cooler, lower luminosity, and longer lived. The large spread in O
is particularly interesting. Given that its
high abundance causes it to play a role second only to Fe in stellar
opacity, stellar models with large
variations in O abundances might be expected to show a significant impact on
stellar evolution and observable quantities. 

We explore whether the observed level of abundance ratio variations cause significant changes in the evolution of stars and therefore their habitable zones. Section~\ref{s:sims} describes the simulations used, Section~\ref{s:results} presents the quantitative changes in the stellar evolution and habitable zones, and Section~\ref{s:discussion} examines implications of the work. 

\section{Simulations \label{s:sims}}
For this work one solar mass stars were simulated using the stellar evolution code TYCHO \citep{young_2005_aa}. TYCHO is a 1D stellar evolution code with a hydrodynamic formulation of the
stellar evolution equations.  It uses OPAL opacities \citep{iglesias_1996_aa,alexander_1994_aa,rogers_2002_aa}, a combined OPAL and Timmes equation of state (HELMHOLTZ) \citep{timmes_1999_aa,rogers_2002_aa}, gravitational settling (diffusion) \citep{thoul_1994_aa}, general relativistic gravity, 
automatic rezoning, and an adaptable nuclear reaction network with a sparse solver. A 177 element network terminating at $^{74}$Ge is used
throughout the evolution. The network uses the latest REACLIB rates 
\citep{rauscher_2000_aa, angulo_1999_aa,iliadis_2001_aa,wiescher_2006_aa}, weak rates from \citet{LMP00}, and screening from
\citet{graboske_1973_aa}. Neutrino cooling from plasma processes and the Urca
process is included. Mass loss is included but is trivial for a 1\msol main sequence star. (Heightened early mass loss seen in some young stars \citep{wood05} is not included.) It incorporates a description of turbulent convection \citep{MA07_conv, AMY09, arnett_2010_aa, AM11}. 
It has no free convective parameters to adjust, unlike mixing-length theory.

A solar composition from \citet{lo09} was adjusted to the mean abundance ratios of the AAT sample for the elements observed, with all other elements being maintained at solar values for the standard composition. The adjusted values are listed in Table~\ref{tab1}. The intrinsic physical spreads $u_{intrinsic}$  remaining when  observational errors are accounted for \citep{desilva06, so09} in the element to iron ratios are the basis for the variant models. The methodology is described fully in a forthcoming paper \citep{pagano12}, but the relevant values are quoted here. 
Table~\ref{tab1} gives the enhancement/depletion in X/Fe for each element. In each model one element was varied to the 3$u_{intrinsic}$ extremes of the distribution, giving one standard composition model and two models for each of the elements C, O, Na, Mg, Al, Si, Ca, Ti. Of the set of elements in the observations mentioned in Section~\ref{s:intro}, we do not model variations in Cr and Ni because of their very small ranges or the r- and s-process because of their very low abundance. We do include Si in spite of its small variation because it has a relatively high abundance. All other elements are held constant.  

Determinations of O abundance are notoriously sensitive to non-LTE effects \citep{gas07}, which may introduce random unphysical errors in the O abundance measurement not accounted for in the quoted observational errors, so we also examine a smaller range of O variation. Instead of choosing a random value, we use the  $u_{intrinsic}$ for a sample of 40 "solar twins" from \citet{ram09}. Qualifying as solar twins requires observable parameters quantitatively close to the sun. In order for a star to match the sun closely in physical observables, it is necessary for it to have a composition close to solar. Using $u_{intrinsic}$ for the solar twin sample therefore provides a very conservative estimate of [O/Fe] variation. 

Custom opacity tables for each composition were obtained from the OPAL project \citep{iglesias_1996_aa}. All models were run until core hydrogen exhaustion. The models provide high time resolution histories of luminosity and effective temperature. Translation of these quantities into the extent of a habitable zone is anything but straightforward. Habitability depends on, among other factors, atmospheric composition, albedo, density structure and therefore planet mass, cloud cover, assumptions about water loss, geophysical exchange with the atmosphere, and assumptions about runaway greenhouse effects. For this work we chose a reasonably conservative prescription for an Earth mass planet that avoids the most extreme predictions for boundaries. We chose to estimate the inner and outer radii of the habitable zone using the prescription of \citet{kasting93, selsis07}. The radii depend upon stellar $T_{eff}$ and luminosity, with
\begin{equation}
l_{in} = (l_{in\odot} - a_{in}T_{*} - b_{in}T_{*}^2)\left(\frac{L}{L_{\odot}}\right )^{\frac{1}{2}}
\end{equation}
and
\begin{equation}
l_{out} = (l_{out\odot} - a_{out}T_{*} - b_{out}T_{*}^2)\left(\frac{L}{L_{\odot}}\right )^{\frac{1}{2}}
\end{equation}

 where $a_{in} = 2.7619 \times 10^{-5}$, $b_{in} = 3.8095 \times 10^{-9}$, $a_{out} = 1.3786 \times 10^{-4}$, $b_{out} = 1.4286 \times 10^{-9}$, and $T_{*} = T_{eff}-5700$. For a 0\% cloud cover atmosphere and assuming an inner boundary set by the water loss limit, $l_{in\odot} = 0.95$AU. For a 50\% cloud cover atmosphere, $l_{out\odot} = 1.95$AU.  We choose the larger inner boundary in order to give a 1 AU orbit for a solar composition a habitable lifetime of 5.5-6 Gy. With this type of treatment, the habitable zone edges can be easily scaled from the stellar quantities for different assumptions, and the relative differences between the stellar models are robust.

\section{Results \label{s:results}} 
\begin{figure}[h]
\begin{center}$
\begin{array}{cc}
\includegraphics[width=2.75in]{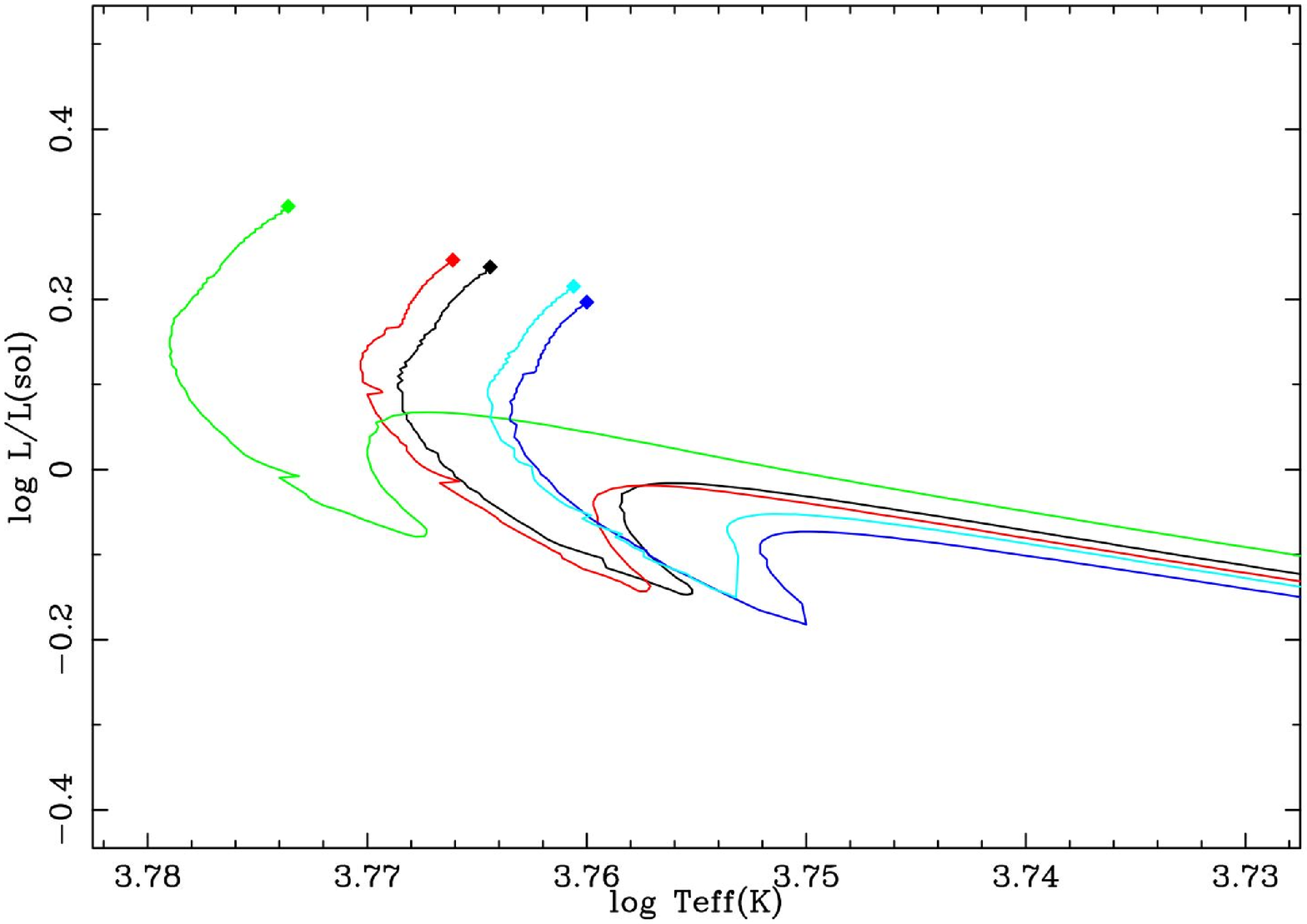} &
\includegraphics[width=2.75in]{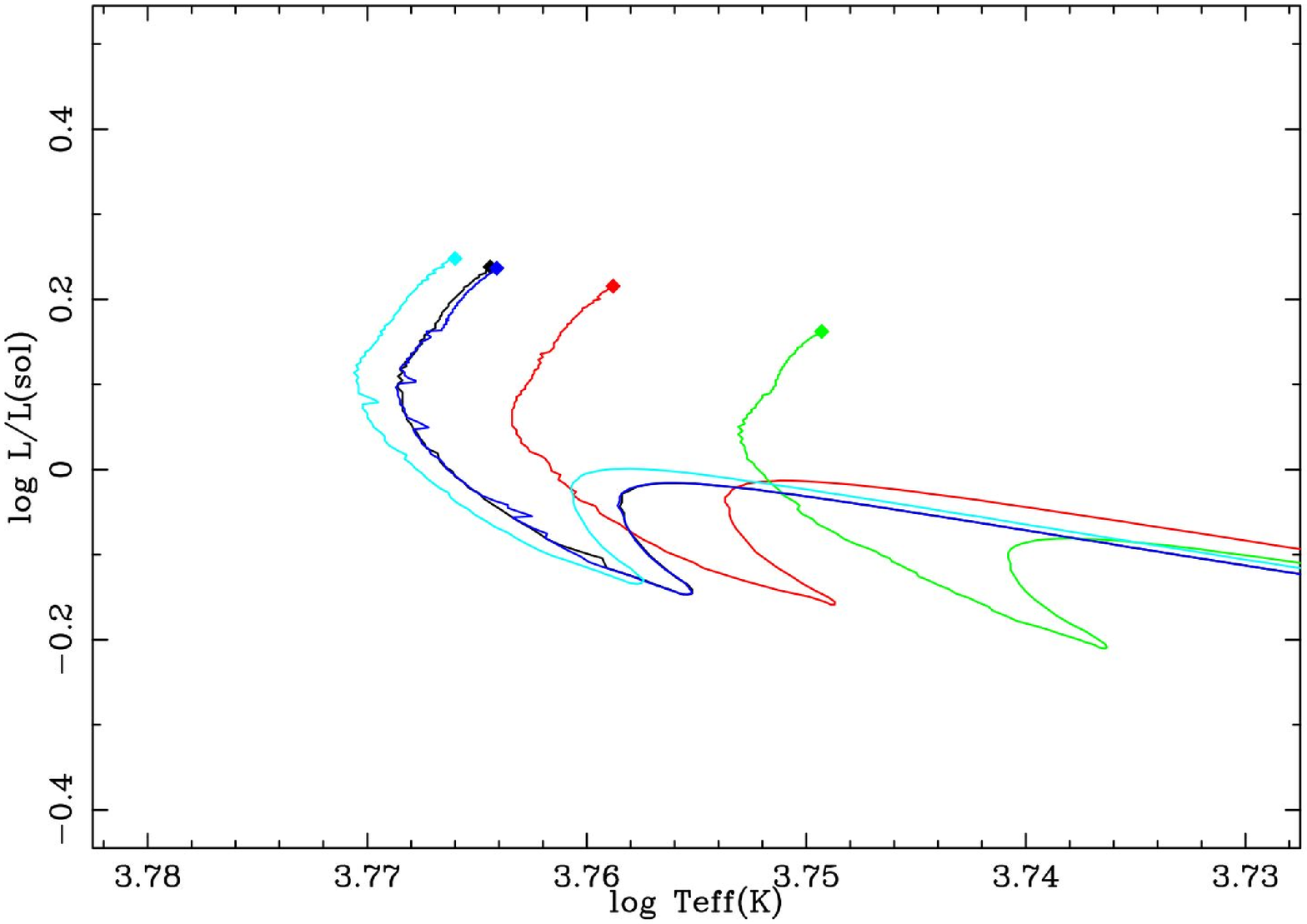} \\
\end{array}$
\end{center}
\caption{Evolutionary tracks for depleted (left) and enhanced (right) abundance ratios. Black is the standard composition of \citet{pagano12}. Na (blue), Mg (cyan), C (red), and O (green) are each varied according to Table~\ref{tab1} while the rest of the metals are held constant. The effects of abundance enhancements are consistently larger than those of depletions, driving stars to lower $T_{eff}$ and luminosity. O causes the most pronounced change.  Kinks in the tracks are slight variations in the models' envelope solutions and represent a temperature variation of order 5K. \label{hrfig}}
\end{figure}

The impact of abundance variations for several elements are large enough to be easily seen in the HR diagram, in which the evolutionary tracks of the stellar models are plotted in $\log{L/L_{\odot}}$ versus effective temperature $T_{eff}$(K). Tracks are shown in Figure~\ref{hrfig} for enhancements (left) and depletions (right) in C, O, Na, Mg.  In all cases enhanced abundance ratios cause larger changes than depletions. C, Na, and Mg have small but noticeable effects. 
C has a high abundance but relatively few electron transitions and low ionization potential. Mg and Na have lower abundances but higher opacity per gram than C, resulting in a similar degree of shift in the tracks. Si has less impact due to its smaller range of variation, and Al, Ca, and Ti have very small effects due to their small abundances and can be neglected for our purposes. Although the main sequence tracks are similar in morphology, substantial differences arise on the pre-main sequence (pre-MS). This is because the changed C and O abundances affect the energy release and therefore transient convective core size, during the partial CN cycle burning that occurs during the pre-MS phase. This stage is, however, too short-lived to be relevant to habitability. 

\begin{figure}[h]
\begin{center}
\includegraphics[width=6in]{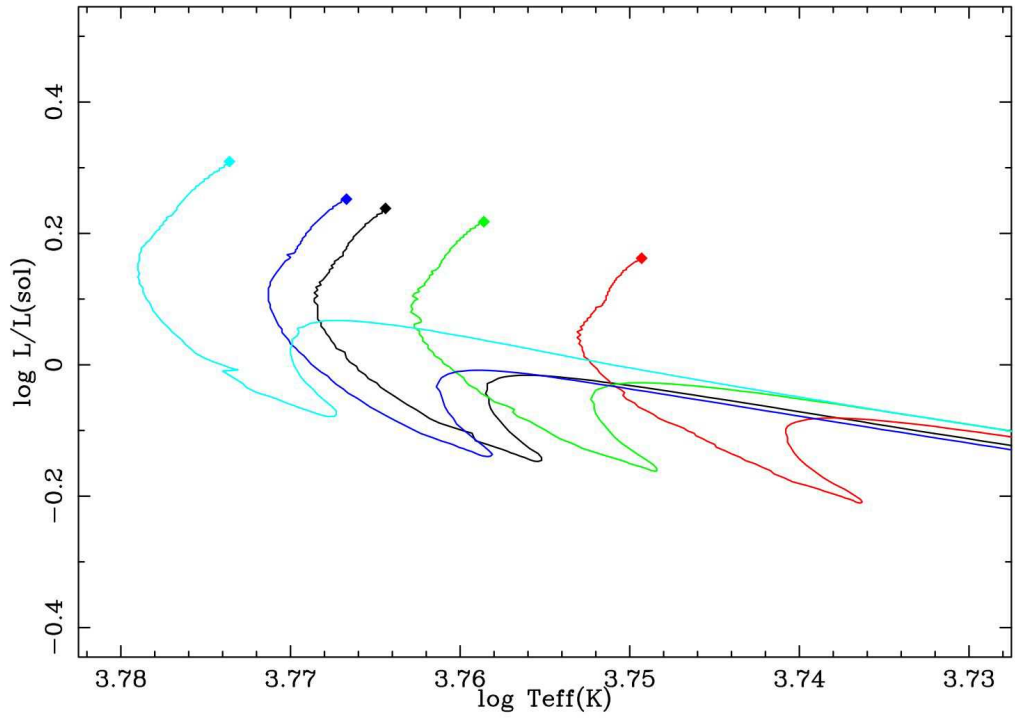}
\end{center}
\caption{ Evolutionary tracks for Standard (black), enhanced O (cyan), moderately enhanced O (blue), moderately depleted O (green), and depleted O (red). Similarly to the tracks for other elements, luminosity and temperature shifts are smaller for depletions than enhancements. \label{hrofig}}
\end{figure}

The largest changes, unsurprisingly, arise from variation in O. 
Figure~\ref{hrofig} shows the evolutionary tracks for the Standard, enhanced, moderately enhanced and depleted, and depleted O compositions. As with the other elements, enhancements result in larger changes than depletions due to the larger absolute changes in mass fraction of the element.  For this letter we will concentrate our analysis on the adjusted O abundance ratios since the evolutionary tracks for O represent the extremes of variation.

Both the luminosity and $T_{eff}$ of enriched compositions are systematically lower at a given age.  The peak temperatures reached on the main sequence are $\log T_{eff} = 3.753$ for enhanced [O/Fe] and $\log T_{eff} = 3.78$ for low [O/Fe] ($\log T_{eff} = 3.762$ for the sun). This gives a temperature difference of $\sim$360K, or a 6\% deviation from solar effective temperature. The temperature range for the moderate O cases is 135K, or 2.3\%. 

\begin{figure}[h]
\begin{center}
\includegraphics[width=6.0in]{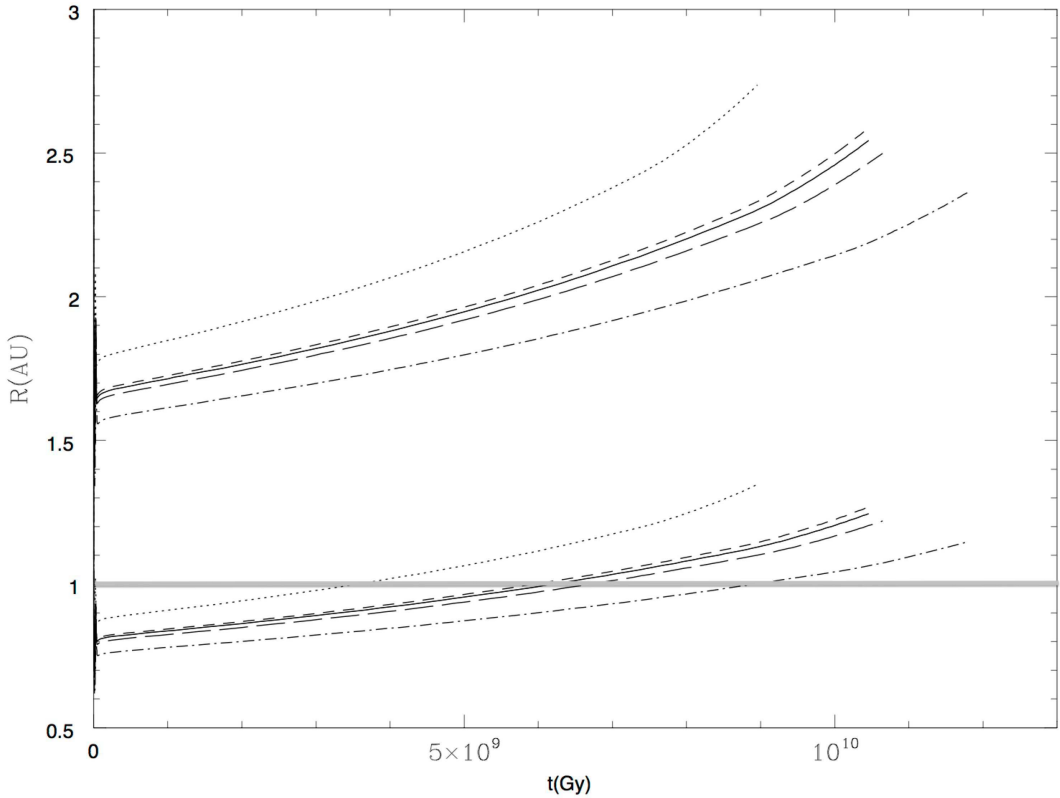} 
\end{center}
\caption{Inner and outer edges of the habitable zone \citep{selsis07} versus time for O. Habitable zones are shown for models with depleted (dotted),moderately depleted (short dash), standard (solid), moderately enhanced (long dash) and enhanced (dot-dash) O abundance ratios. Depletions systematically move the habitable zones outward due primarily to increased luminosity and secondarily to increased $T_{eff}$. The difference in location increases markedly with age. The most dramatic effect results from the substantial difference in main sequence lifetimes. \label{hzfig}}
\end{figure}

The time evolutions show the most profound effect to be on the pace of the evolution. The main sequence turnoff for the low [O/Fe] model occurs at an age of 9Gy. The standard composition (which is somewhat more O-rich than solar) has a turn-off age of 10.5Gy, and the O-rich model turns off at 12Gy at lower $\log L$ and $T_{eff}$. A 33\% increase in stellar lifetime is interesting in itself, but the effects on habitability require further analysis. Orbits with semi-major axes near the edges of habitable zones will not remain there for the entire lifetime of the star. Earth itself is in this predicament, with many models predicting only about a billion years more of habitability. Figure~\ref{hzfig} shows the inner and outer edges of habitable zones for the high, low, and standard [O/Fe] as a function of time, calculated using the \citet{selsis07} prescription. The gray bar marks a 1AU orbit. We consider several properties of habitable orbits. First, we find the innermost and outermost continuously habitable orbits and innermost and outermost orbits habitable for 4Gy. An Earth-centric measure of habitable zone persistence is the habitable lifetime of a 1AU orbit.  The radial extent of the habitable zone varies substantially over the life of the star as well, increasing monotonically. The evolution of the outer edge of the habitable zone in particular is steepest and largest in absolute terms for the O depleted model. These parameters are provided in Table~\ref{tab2}.

\section{discussion \label{s:discussion}}
Potential for habitability will be an important consideration when allocating resources for the very difficult direct observations of exoplanets \citep{tess12, vonp11, rauer11, kt09}. A first cut of candidates can be made on the {\it current} habitability of a planet estimated directly from the measured properties of the host star. The current habitability of a planet is not the only consideration in finding Earth-like worlds, however. If the ultimate goal is the discovery of life, the evolution of habitable zones must be taken into account. Life did not modify the Earth's atmospheric composition at levels that might be detectable by near-future astronomical observations until  about a billion years after its formation. Complex life on Earth developed during the late Ediacaran period, 635 - 542 Mya, some 4Gy after its formation. For the holy grail of Earth-like complex life, if Earth is at all representative, we should look for planets that have spent several billion years inside their habitable zones.

For a star otherwise identical to the sun in mass and Fe abundance at the lower end of the range of [O/Fe] observed in planet host candidates, an Earth-like planet at 1AU from the star would not (given the assumptions about planetary atmospheres herein) remain habitable long enough for  complex life to have developed at the pace that occurred on Earth. The effect is obviously dramatic for the extremes of O variation, but there is even an effect that could be important to the less extreme systems with planets on the edges of their nominal habitable zones. The 1 AU habitability lifetime differs by nearly a billion years between the moderately enhanced and depleted cases. This is the variability deduced for a set of stars deliberately chosen to be maximally sun-like. The magnitudes of the effect for the observed ranges of [C/Fe], [Na/Fe], and [Mg/Fe] are hundreds of millions to a billion years.

Any orbit that is not continuously habitable varies in its habitable lifetime by billions of years between stars of different abundance ratios. If we limit ourselves to systems that are continuously habitable for the entire main sequence lifetime of the star for all [O/Fe] we limit ourselves exceedingly. For this atmospheric prescription the only orbits habitable for all of the models with the observed range of [O/Fe] at solar [Fe/H] extend from 1.35AU to 1.55AU. Ranges are similarly narrow though shifted in radius for other [Fe/H]. For any given single composition the range is much larger, increasing for lower abundance ratios.  We also find the orbits that should have at least 4Gy of continuous habitability. This broadens the range of allowed orbits considerably. High [O/Fe] models have the widest range of 4Gy habitable orbits by a small margin due to their longer overall main sequence lifetime and shallower rate of change of the zone boundaries. A point that is not commonly considered is that the majority of these orbits are habitable during the latter part of the star's lifetime, meaning that many of the planets that may be favorable to the development of life that we discover will have only begun their habitable lifetimes. Also, while these ranges encompass the majority of nearby planet host candidates, there are even more extreme outlier systems. The stars HD53705 ([O/Fe] = 4) and HD136352 ([C/Fe] also $\sim$4)have a substantially larger range of orbits that remain habitable for a multi-Gyr window with such large enhancements. 

Conversely, this work also reiterates the much greater difficulty in determining the extent of a star's habitable zone due to planetary conditions. Our discussion is limited to the effects on the star. The pace of a star's evolution can cause a very large change in the habitable lifetime of an orbit even if the radial distance of the habitable zone from the star changes by only a small amount. At 6.1Gy the habitability of the 1AU orbit ends for the standard composition. The inner edge of the habitable zone for the moderately enhanced O case at 6.1Gy is at 0.97AU. This 3\% difference results in an additional 0.6Gy of habitability due to the slower increase in luminosity of the more O-rich composition. A very minor change in the model parameters chosen even from the single atmospheric prescription we examine would change the habitable zone radius more than this. This also assumes that prescriptions for planetary habitability based on assumptions about the evolution of planetary atmospheres are appropriate for different planetary compositions. \citet{bond08} show that the mineralogy of solid bodies arising from protoplanetary disks of non-solar abundance ratios can be dramatically different from the solar system. The composition of planetary atmospheres and the geological cycles that modify and maintain them can be expected to vary similarly. 

The viability of exoplanets at the edges of their habitable zones depends directly on the abundances of all the common elements.  The stellar characteristics that determine the size and persistence of habitable zones, completely apart from properties of the planets themselves, change at an important level. Proper characterization of a system requires more data than just [Fe/H]. A concerted effort to obtain high quality spectral observations of all planet hosts of astrobiological interest would be valuable.

{\bf Acknowledgements:}
We would like to thank Eric Mamajek and Dave Arnett for useful discussions, Mark Richardson, Matt Mechtley, and Vahid Golkhu for abundance tables, and the referee for most excellent suggestions for improving the paper. This work was supported in part by a NASA Astrobiology Institute grant.

\begin{deluxetable}{lccc}
\tablewidth{0pt}
\tablecaption{Abundance Enhancements/Depletions Relative to Standard Composition\label{tab1}}
\tablehead{
  \colhead{Element}

& \colhead{Depletion}  
& \colhead{Enhancement}
& \colhead{[X/H]$_{standard}$\tablenotemark{a}}
}

\startdata
C  &  0.48 & 2.09 & 0.173 \\
O  &  0.44 & 2.28  & 0.051\\
Moderate O\tablenotemark{b} & 0.67 & 1.48 & \\
Na  &   0.58 & 1.74 & 0.068 \\
Mg  & 0.43 &  2.32 & -0.029\\
Al  &  0.55 & 1.81 & -0.124 \\
Si  &    0.71 & 1.41& 0.078 \\
Ca  &  0.54 & 1.86& 0.027\\
Ti &  0.40 & 2.49 & 0.072\\

\enddata
\tablenotetext{a}{Standard composition values relative to solar}
\tablenotetext{b}{Uses the smaller range of O variation observed in twins (see Section~\ref{s:sims}).}
\end{deluxetable}
\begin{deluxetable}{lccccc}
\tablewidth{0pt}
\tablecaption{Habitable Zone Extent \label{tab2}}
\tablehead{
  \colhead{Composition}

& \colhead{Continuously}  
& \colhead{4Gy Habitable}
& \colhead{Width at}
& \colhead{Width at H} 
& \colhead{Habitable Lifetime} \\
 \colhead{}
& \colhead{Habitable}  
& \colhead{}
& \colhead{ZAMS}
& \colhead{Exhaustion} 
& \colhead{of 1AU Orbit} \\
\colhead{}
& \colhead{AU}  
& \colhead{AU}
& \colhead{AU}
& \colhead{AU} 
& \colhead{Gy}
}

\startdata
High O  &  1.15 - 1.55 & 0.84 - 2.11 & 1.3 & 1.7 & 9.0 \\
Medium High O & 1.22 - 1.64 & 0.9 - 2.04 & 1.34 & 1.79 & 6.7 \\
Standard  &  1.25 - 1.65 & 0.92 - 2.06  & 1.35 & 1.8 & 6.2 \\
Medium Low O & 1.25 - 1.67 & 0.94 - 2.19 & 1.35 & 1.8 & 5.8 \\
Low O  &   1.35 - 2.12 & 1.03 - 2.2 & 1.41 & 1.89 &  3.5 \\

\enddata
\end{deluxetable}

\end{document}